# Deep Learning in Memristive Nanowire Networks

Jack D. Kendall, Ross D. Pantone, and Juan C. Nino

*Abstract*— Analog crossbar architectures for accelerating neural network training and inference have made tremendous progress over the past several years. These architectures are ideal for dense layers with fewer than roughly a thousand neurons. However, for large sparse layers, crossbar architectures are highly inefficient. A new hardware architecture, dubbed the MN3 (Memristive Nanowire Neural Network), was recently described as an efficient architecture for simulating very wide, sparse neural network layers, on the order of millions of neurons per layer. The MN3 utilizes a high-density memristive nanowire mesh to efficiently connect large numbers of silicon neurons with modifiable weights. Here, in order to explore the MN3's ability to function as a deep neural network, we describe one algorithm for training deep MN3 models and benchmark simulations of the architecture on two deep learning tasks. We utilize a simple piecewise linear memristor model, since we seek to demonstrate that training is, in principle, possible for randomized nanowire architectures. In future work, we intend on utilizing more realistic memristor models, and we will adapt the presented algorithm appropriately. We show that the MN3 is capable of performing composition, gradient propagation, and weight updates, which together allow it to function as a deep neural network. We show that a simulated multilayer perceptron (MLP), built from MN3 networks, can obtain a 1.61% error rate on the popular MNIST dataset, comparable to equivalently sized software-based networks. This work represents, to the authors knowledge, the first randomized nanowire architecture capable of reproducing the backpropagation algorithm.

## I. Introduction

CURRENTLY, deep learning architectures are heavily constrained in terms of the number of neurons that can be simulated in a typical fully-connected layer. This is primarily due to the quadratic scaling of compute time for vector-matrix multiplication with increasing layer size [1]. This scaling sets the practical upper limit of fully-connected layer sizes for deep neural networks to around 20,000 × 20,000 neurons. By contrast, other common operations in deep learning, such as computing activation functions and performing normalization, scale linearly with layer size. It is precisely because of the quadratic scaling of vector-matrix multiplication with respect to layer size that narrow and deep architectures are more efficient to implement on current hardware than wide architectures. To better visualize the situation, it is useful to note that doubling the number of neurons in a network of fully connected layers in depth doubles the number of operations, while doubling the number of neurons in width quadruples the number of operations. Nonetheless, recent research shows that the most powerful networks are both wide and deep [2, 3, 4, 53].

It is commonly understood that to exploit the benefits of wide layers in deep learning, two strategies can be employed. The first is to accelerate the number of operations per second (OPS) of current GPU-like architectures. Within this general strategy, reducing precision, increasing the number of multiprocessors per chip, and improving data movement are promising approaches being explored by a number of groups [5, 6, 7]. The second strategy is to leverage computing architectures which do not have $O(n^2)$ time scaling with respect to vector-matrix multiplication.

This work was supported by Rain Neuromorphics, Inc. The authors have a financial interest in the proposed technology presented in this paper. J. D. Kendall and J. C. Nino are co-owners and R. D. Pantone is an employee of Rain Neuromorphics, Inc.

J. C. Nino acknowledges the support of the US National Science Foundation under Grant No. ECCS-1709641.

R. D. Pantone and J. D. Kendall are with Rain Neuromorphics, Inc., Redwood City, CA, USA. (e-mail: ross@rain-neuromorphics.com; jack@rain-neuromorphics.com).

J. C. Nino is with the Department of Materials Science and Engineering, University of Florida, Gainesville, FL, USA. (e-mail: jnino@mse.ufl.edu).

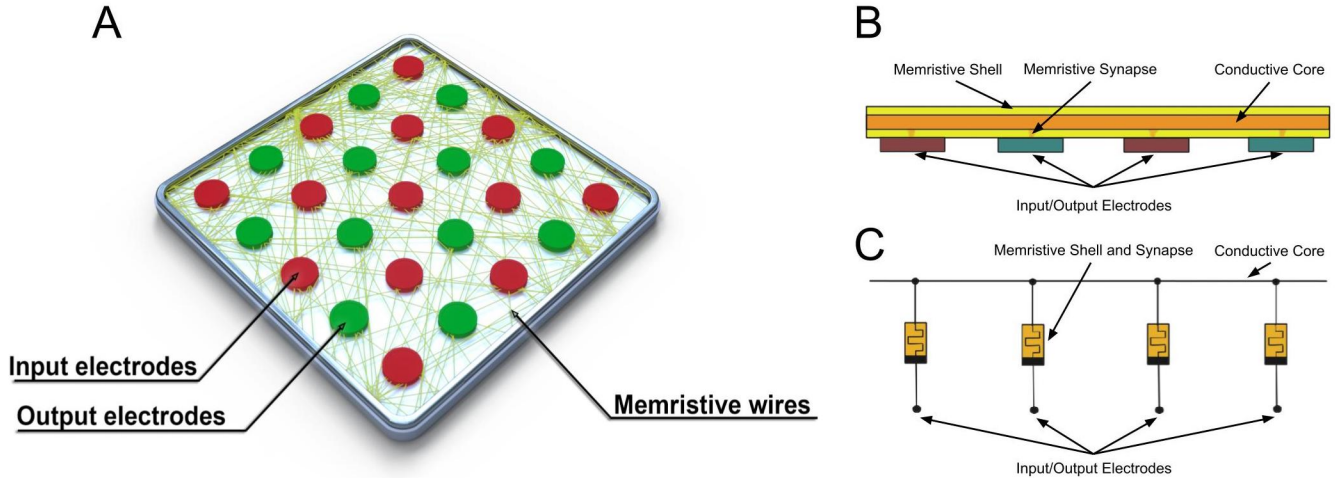

Figure 1. (a) An example MN3 simulated using the straight wire model in [27], where each red-green electrode pair represents a neuron of a standard neural network and each yellow line represents a stochastically deposited nanowire. Note that input and output electrodes may be designated arbitrarily. These electrodes are connected with a nanowire mesh layer. (b) Intersecting nanowires and electrodes form memristive synapses. (c) The equivalent circuit diagram of Fig. 1b.

When exploring options within this second strategy, one finds that the most well-known architectures that are able to circumvent the quadratic scaling of vector-matrix multiplication with respect to time are in-memory computing architectures known as dot-product engines [8, 9]. These architectures operate by exploiting the natural vector-matrix multiplication that occurs as a result of Ohm's law. A vector of voltages is multiplied by a matrix of conductances to produce a vector of currents in a single computational step. Loading and reading of data are linear in complexity and therefore, the entire computation can be performed with $O(n)$ complexity with respect to time [8, 9]. Essentially, this approach removes matrix multiplication as the fundamental bottleneck for neural network architectures and would allow networks to scale to much larger sizes.

Unfortunately, for the purposes of exploiting this advantage, nearly all current dot product engine computing architectures are based on some variant of the crossbar array [8, 9, 10, 11, 12]. These crossbar arrays are fully connected, meaning every input neuron is connected to every output neuron in the array. This connectivity results in quadratic scaling with respect to the spatial resources of the chip. As the layer size increases, the required die size rapidly exceeds what is possible to fabricate reliably. As a result, current crossbars are limited to fewer than a thousand neurons in size, which restricts their use in very wide networks [13]. Network-on-chip architectures such as IBM's TrueNorth are able to tile small crossbars into a large array and connect those using serial busses; this approach, however, results in a similar serial communication bottleneck as in current GPUs [14]. Other approaches such as mesh networks have also been proposed and implemented [15, 16, 17], but they carry their own constraints. Specifically, meshes have long latencies as the average minimal path scales by $O(\sqrt{n})$ and deadlock-free communication cannot be guaranteed [18, 19, 20].

It is therefore fairly clear that in order to implement very large layer sizes in a fully parallel in-memory computing architecture, efficient sparsity of connections is needed. For example, in biological neural networks, small-world and scale-free sparsity minimizes the number of connections necessary, while keeping the average path length between two randomly selected neurons very low. The average minimal path length scales by $O(\log n)$, maintaining the network's ability to efficiently process information [21, 22]. At the same time, neuron and synapse densities are very high. The result is a massive network with high degrees of structured sparsity and randomness [23, 24].

TABLE I

The left column provides a list of notable architectures. The right column contains the corresponding largest layer-to-layer transformations. The MN3 contains three to four orders of magnitude more neurons per layer than any well-known deep architecture to the authors' knowledge. We extrapolate our estimate from [27] assuming the use of a full-size 600 mm$^2$ reticle for fabrication. Note that any MN3 transformation with layer sizes that sum up to 24,000,000 is possible.

| Architecture | Largest layer-to-layer transformation (number of neurons) |
|---|---|
| AlexNet [28] | 9,216 × 4,096 |
| VGG Net [29] | 25,088 × 4,096 |
| ResNet [30] | 2,048 × 1,000 |
| Transformer [31] | 1,024 × 4,096 |
| LSTM for Language Modeling [32] | 8,192 × 2,048 |
| Deep Speech 2 [33] | 2,560 × 2,560 |
| Block-Sparse LSTM [34] | 18,432 × 18,432 |
| T5-11B [53] | 65,536 × 1,024 |
| MN3 [27] | 12,000,000 × 12,000,000 |

We have previously described a new hardware architecture dubbed the MN3 (memristive nanowire neural network) that was designed with these biological neural network principles in mind and represents an in-memory computing architecture with exceptional scaling properties [25, 26, 27]. We showed that the MN3 exhibits sparse, small-world connectivity, which allows the network to scale to very large layer sizes without a quadratic increase in the number of synapses, while keeping the network well-connected, i.e. maintaining a low average path length and high clustering coefficient [27].

In the following sections, we further explore the MN3 architecture and assess its neuron and synapse densities, and most importantly, we demonstrate that the network can be trained using backpropagation and a variant of stochastic gradient descent (SGD). The main implication being that layers made from these networks can be linked together and trained as a deep neural network using standard optimization techniques.

We first describe the forward pass of the MN3 in analog terminology. Next, we describe the weight update mechanism in detail, including sources of error and potential limitations. We then present results on detailed simulations of various MN3 networks and their performance on standard deep learning benchmarks, with an emphasis on comparisons to standard networks trained using gradient descent. Finally, we discuss areas of future focus and conclude with a few remarks on scalability and the overall implications of our results.

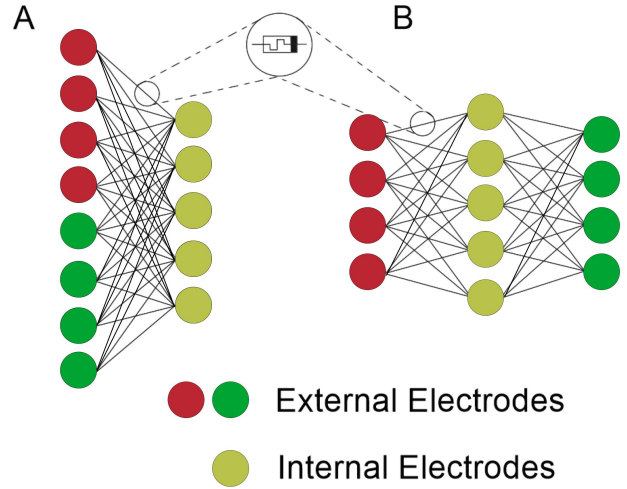

Figure 2. (a) Bipartite memristive network, where blue nodes represent input electrodes, yellow nodes represent nanowire cores, and red nodes represent output electrodes. (b) Alternate feedforward form with inputs and outputs partitioned.

## II. ARCHITECTURE (REVISITED)

The MN3 architecture comprises two parts: a silicon chip housing artificial neurons and I/O circuitry, and a nanowire layer placed on the surface of the silicon which serves as the interconnect and synapse network for the neurons. Each nanowire has a core-shell structure, with a metal core to allow for signal transmission over long distances and a memristive shell, which creates trainable synapses at the interfaces between each electrode and nanowire. The resulting MN3 architecture is illustrated in Fig. 1.

By offloading the wiring and synapses to the nanowire layer, a large amount of on-chip resources can be freed and dedicated to the neurons. At the same time, the 2.5D stacking of the nanowires enables high synapse densities. Assuming 100 nm diameter wires in a 10 wire-thick mat with 25 μm$^2$ neurons, this allows for a neuron density of $4 \times 10^6$ neurons per cm$^2$ and a synapse density of $4 \times 10^8$ synapses per cm$^2$ [27]. In a full-size 600 mm$^2$ reticle for CMOS fabrication, this would allow for 24M neurons to be integrated on one die with one-step matrix multiplication, and 100 synapses per neuron. Note that the number of synapses per neuron can be traded for neuron density. So, if 1,000 synapses per neuron are required, we can simply reduce the neuron count by a factor of 10. This is done by aggregating multiple electrodes into one neuron. For comparison, Table I provides the largest layers of several conventional deep architectures by number of neurons.

Although the MN3 is not a fully connected network, the connectivity pattern of the randomly aligned memristive nanowires exhibits small-world properties [27]. This means that despite the high degree of sparsity in the network (i.e. not every neuron is directly connected to every other neuron), the average path length connecting two neurons is very short, allowing information to be propagated efficiently through the network. This small-world connectivity resembles the connectivity patterns observed in biological brains [23].

The MN3 also possesses another important network property: bipartite connectivity. This property allows for the individual memristors in the network to be collectively adjusted according to an arbitrary cost function (described in detail in the following sections). Fig. 2 depicts the bipartite connectivity of the MN3 architecture. The first set of nodes is the set of external electrodes, and the second set of nodes is the set of nanowire cores. The external electrodes have no direct connections to other external electrodes in the CMOS layer, and the wire cores are effectively insulated from one another due to the very high interface resistance of the junction between two memristive oxide shells. This means that the only connections present are between external electrodes and nanowire cores.

By partitioning the external electrodes (i.e. the electrodes connected to the CMOS layer) into inputs and outputs, a two-layer neural network can be formed. Here, the input and output neurons are connected by a linear hidden layer (i.e. the nanowire cores). Since the output currents can be read out digitally, we can apply a nonlinear activation function such as ReLU, tanh, or softmax. By connecting the outputs of one MN3 to the inputs of another after this nonlinear activation function, multiple of these networks can be chained together to form a deep neural network. We can also apply other standard deep learning functions such as layer normalization, pooling, and dropout between MN3 transformations.

### III. Algorithm Overview

The proposed architecture functions as a dot-product engine for vector-matrix multiplication: a vector of voltages is multiplied by a matrix of conductances to produce a vector of currents. For a brief overview of how these architectures operate, see [8, 9].

We can envision the graph in Fig. 2b as a two-layer neural network. If we require more hidden layers, several such networks can be linked together with nonlinear activation functions between them. For simplicity, we will analyze a two-layer network. In order to determine how the network behaves, we can use Kirchhoff's laws to determine the voltage drops across the memristors. Once we have the voltage drops and currents through the network, we can apply an appropriate memristor model to determine how the conductances of these memristors change over time.

First, the $N_1$ inputs to the neural network are presented as voltages between $-V_T / 2$ and $V_T / 2$, where $V_T$ is the threshold voltage of the memristor. This avoids changing any memristor conductances. Next, we assume that the $N_2$ outputs of the neural network are read off as currents at the output nodes, where the voltages are kept at ground. Since the total current $I_j^{\text{int}}$ into any internal node $j$ must sum to zero (Kirchhoff's current law), we have the following equation:

$$I_j^{\text{int}} = \sum_{i=1}^{N} G_{ij}\left(V_i^{\text{ext}} - V_j^{\text{int}}\right) = 0, \quad (1)$$

where $G_{ij}$ is the instantaneous conductance of the memristor connecting external node $i$ to internal node $j$, $V_i^{\text{ext}}$ is the external voltage at node $i$, $V_j^{\text{int}}$ is the internal voltage at node $j$, and the sum runs over all $N = N_1 + N_2$ external nodes, i.e., the input *and* output nodes.

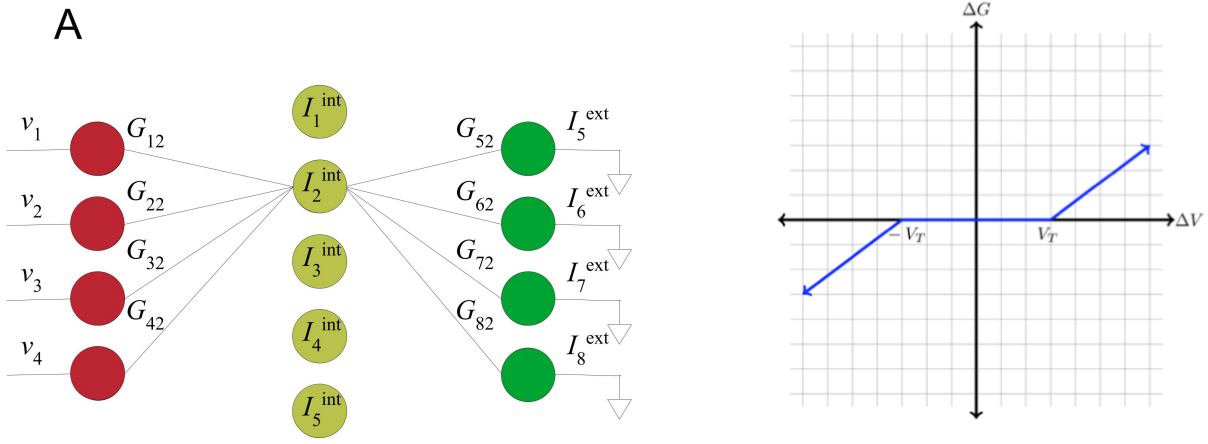

**Fig. 3.** (a) Example feedforward structure for all external connections with one internal node (namely, the second node from top) with input voltages $v_1, \ldots, v_4$, grounded output voltages ($v_5 = \ldots = v_8 = 0$ V), conductances $G_{12}, \ldots, G_{82}$, internal currents $I_1^{int}, \ldots, I_5^{int}$, and external currents $I_5^{ext}, \ldots, I_8^{ext}$ to be read in digitally. For the external currents, we start at index 5 as we intentionally omit the first four external currents from the image. Note that $I_1^{int} = \ldots = I_5^{int} = 0$ by Kirchhoff's law. (b) Linear model of a memristor with symmetric threshold $V_T$. Note that memristor switching will only occur if the voltage across the memristor is greater than $|V_T|$.

We can rearrange (1) to solve for the voltage at internal node $j$ as follows. Since the summation is linear and $V_j^{int}$ is not indexed by $i$:

$$\sum_{i=1}^{N} G_{ij} V_i^{ext} = \sum_{i=1}^{N} G_{ij} V_j^{int},$$

$$\sum_{i=1}^{N} G_{ij} V_i^{ext} = V_j^{int} \sum_{i=1}^{N} G_{ij},$$

$$V_j^{int} = \frac{\sum_{i=1}^{N} G_{ij} V_i^{ext}}{\sum_{i=1}^{N} G_{ij}} = \frac{\sum_{i=1}^{N} G_{ij} V_i^{ext}}{G_j}, \quad (2)$$

where we define the normalization factor $G_j = \sum_{i=1}^{N} G_{ij}$, which does not depend on the instantaneous values of the voltages or currents. While this normalization term $G_j$ is not present in standard implementations of neural networks, we find that it does not significantly influence the operation of the network, as shown in section (V). If we further note that all of the output nodes are left at ground ($V_i^{ext} = 0$), the only contributions to the sum are those of the input nodes. Assuming the voltages are between $-V_T/2$ and $V_T/2$, there will be no memristor switching, and we can view the memristor network as a resistor network. Since transformations are linear in a resistor network, we can think of the voltages at the internal nodes as a linear transformation of the input by a matrix of conductances [35].

Similarly, we can solve for the currents at the output node $k$ using the $M$ voltages at the internal nodes. Since the output nodes are all grounded (see Fig. 3a), the current is given by:

$$I_k^{ext} = \sum_{j=1}^{M} G_{kj} V_j^{int}. \quad (3)$$

Again, this is a linear transformation of the hidden nodes by a matrix of conductances. However, since this current is accessible by an external circuit, a further nonlinear transformation can take place at this layer, allowing the network to possess nonlinear activations.

## IV. BACKPROPAGATION AND WEIGHT UPDATE ALGORITHM

Now that we have framed a single MN3 in terms of a two-layer neural network with memristive connections, an appropriate algorithm can be implemented in order to train the network. For this purpose, we describe an approximation to the backpropagation algorithm that is suitable for training the network. As the network size becomes large, the approximate gradients approach the exact gradients. Subsection (A.4) provides details on the nature of this approximation. It is important to note that this approximation is primarily designed for large networks, where the difference between approximate gradients and exact gradients is small.

46Section (V) shows its effectiveness on reasonably sized networks (layers containing ~1,000 neurons).We begin by describing the memristor model we use throughout this paper. The model is inspired by [36], which describes a linear threshold memristor model. This is the simplest model of a memristor with a voltage threshold, below which no resistance change occurs. Again, we recognize the simplicity and idealism of this model. We emphasize that this model is primarily used as a steppingstone for more realistic memristor models. The main difference between [36] and our model is that our model is defined in terms of conductances, rather than resistances. This is for computational simplicity in computing weight (conductance) changes, as it eliminates the need for taking inverses during simulation of the network. The graph of the rate of conductance change vs voltage drop across the memristor is given in Fig. 3b. For a fixed voltage drop $\Delta V$ across the memristor, applied for duration $\Delta t$, the net conductance change $\Delta G$ of the memristor is given by:$$\Delta G = \begin{cases} \beta(\Delta V - V_T)\Delta t & \text{if } \Delta V > V_T \\ \beta(\Delta V + V_T)\Delta t & \text{if } \Delta V < -V_T \\ 0 & \text{if } -V_T < \Delta V < V_T \end{cases} \quad (4)$$
6

Section (V) shows its effectiveness on reasonably sized networks (layers containing ~1,000 neurons).

We begin by describing the memristor model we use throughout this paper. The model is inspired by [36], which describes a linear threshold memristor model. This is the simplest model of a memristor with a voltage threshold, below which no resistance change occurs. Again, we recognize the simplicity and idealism of this model. We emphasize that this model is primarily used as a steppingstone for more realistic memristor models. The main difference between [36] and our model is that our model is defined in terms of conductances, rather than resistances. This is for computational simplicity in computing weight (conductance) changes, as it eliminates the need for taking inverses during simulation of the network. The graph of the rate of conductance change vs voltage drop across the memristor is given in Fig. 3b. For a fixed voltage drop $\Delta V$ across the memristor, applied for duration $\Delta t$, the net conductance change $\Delta G$ of the memristor is given by:

$$\Delta G = \begin{cases} \beta(\Delta V - V_T)\Delta t & \text{if } \Delta V > V_T \\ \beta(\Delta V + V_T)\Delta t & \text{if } \Delta V < -V_T \\ 0 & \text{if } -V_T < \Delta V < V_T \end{cases} \quad (4)$$

The proportionality constant $\beta$ gives the rate of change in conductance for a given voltage drop, and its sign determines the polarity of the memristor (i.e. whether positive voltage drops increase or decrease conductance).

This model is an idealization of real memristors, which have many nontrivial effects such as asymmetric, nonlinear conductance changes, dependence on temperature, and state-dependent dynamics. There is a growing body of literature on handling these nonidealities in neuromorphic hardware [37, 38]. However, for simplicity in deriving and simulating the algorithm, we ignore most of these nonideal effects. While we leave more detailed simulations with nonideal memristors for future work, our preliminary results in this direction are encouraging.

*A. Output Weight Updates*

For the derivation of our algorithm, we take as our starting point the delta rule. The delta rule is a special case of the more general backpropagation algorithm and can be applied to update the output weights of a multi-layer neural network. It gives the correct output weight updates for stochastic gradient descent learning. We assume a softmax output nonlinearity with a cross-entropy loss function, although other nonlinearities and loss functions can be considered. The form of the delta rule is then given by:

$$\Delta W_{kj} = \alpha(y_k - T_k)h_j \quad (5)$$

Here, $W_{kj}$ is the weight connecting output node k to hidden node j, $T_k$ is the target value for output k, $y_k$ is the actual value for output k, $h_j$ is the value of hidden node j, and $\alpha$ is the step size or learning rate.

We can map these variables to the MN3 network as follows: $W_{kj} \rightarrow G_{kj}$, where $G_{kj}$ is the conductance of the memristor connecting internal node (i.e. wire core) j to output node (i.e. external electrode) k; $y_k \rightarrow \sigma(I^{\text{ext}})_k$, which is the softmax nonlinearity at node k applied to the measured output current vector $I^{\text{ext}}$; $h_j \rightarrow V_j^{\text{int}}$, the voltage at internal node j, and $T_k \rightarrow T_k$, the target output at node k. Now, the delta rule in terms of these variables is given by:

$$\Delta G_{kj} = \alpha(\sigma(I^{\text{ext}})_k - T_k)V_j^{\text{int}} \quad (6)$$

Alternatively, one can derive this equation directly from Eq. (3), which is identical to the pre-activation transformation for the final layer of a multi-layer perceptron. The final output can be obtained by applying the softmax nonlinearity to the measured currents, and the gradient of cross-entropy loss with respect to the $G_{kj}$'s will yield the above equation. Note that this ignores the dependence of $V_j^{\text{int}}$ on the normalization term $G_j$, which is the primary approximation we make in this algorithm. We show analytically in A.4 that as the network size grows, the impact of this error term decreases. In practice, we find it has little influence for reasonably sized (>1,000 neuron) networks.

In order to implement the desired weight updates, we must create a correspondence between the conductance changes driven by the memristor dynamics given in Eq. (4) and the desired conductance changes according to Eq. (6). We can accomplish this by exploiting the voltage thresholds of the memristors, their piecewise linearity in both $\Delta V$ and $\Delta t$, and the linearity of the delta rule in both the delta $\delta_k = (\sigma(I^{\text{ext}})_k - T_k)$ and the hidden activation $V_j^{\text{int}}$, as we now show.

Our strategy will be to partition the weight updates $\Delta G_{kj}$ according to the sign of the hidden unit activation $V_j^{\text{int}}$ and the sign of the delta $\delta_k$ they connect. The weight updates can thus be split into four groups: weights for which $V_j^{\text{int}}$ is positive and $\delta_k$ is positive, weights for which $V_j^{\text{int}}$ is positive and $\delta_k$ is negative, etc. We will see that it is possible to update the conductances of each of these four groups of weights in the correct direction and magnitude, while ensuring the other groups of weights remain unchanged.

Our first insight into why this approach might work is to consider what happens when one of the nodes at the output of the network, say node k, which was originally grounded, is pulsed with a voltage equal to the memristor threshold voltage $V_T$ for a duration $\Delta t$. Note that the inputs are at fixed voltages $V_i^{\text{ext}}$, and the hidden nodes are at voltages $V_j^{\text{int}}$, apart from a small perturbation resulting from the pulse itself.

We can immediately see that all the memristors associated with positive $V_j^{\text{int}}$ do not receive any change in conductance. This is because their voltage drop is given by $(V_T - V_j^{\text{int}})$, which is, by definition, less than $V_T$ for positive $V_j^{\text{int}}$. Therefore, only nodes with negative $V_j^{\text{int}}$ receive a conductance change. It is also easy to see that the magnitude of the conductance change will be proportional to $V_j^{\text{int}}$, given that the voltage drop across the memristor is $V_j^{\text{int}}$ past the threshold $V_T$. From Eq. (4), we have:

$$\Delta G_{kj} = \beta(\Delta V - V_T)\Delta t = \beta\left((V_T - V_j^{\text{int}}) - V_T\right)\Delta t = \beta(-V_j^{\text{int}})\Delta t$$

Since $V_j^{\text{int}}$ is negative, that implies $-V_j^{\text{int}}$ is positive, so the sign of the update is determined by the sign of $\beta$. If we apply the pulse for a duration $\Delta t$ proportional to the absolute value of the delta $|\delta_k|$ of node k, we arrive at:

$$\Delta G_{kj} = \beta\gamma|\sigma(I^{\text{ext}})_k - T_k|(-V_j^{\text{int}}) \qquad (7),$$

where $\gamma$ is the constant of proportionality relating $\Delta t$ to $\delta_k$: $\Delta t = \gamma|\sigma(I^{\text{ext}})_k - T_k|$. Now, the sign of the correct weight update according to the delta rule is given by the product of the signs of $\delta_k$ and $V_j^{\text{int}}$. Since we know $V_j^{\text{int}}$ is negative, the sign of the correct update is given by the opposite sign of $\delta_k$. If $\delta_k$ is negative, the update should be positive (i.e. $G_{kj}$ should increase in conductance), and vice versa. Let us fix $\beta$ to be positive, i.e. the polarity of the memristors is such that a positive voltage drop from external electrode to internal wire core increases the conductance, as shown in Fig. 3b. This means that the updates in Eq. (7) are positive, which corresponds to the correct update magnitude and direction for negative $\delta_k$.

Therefore, the process of applying a positive pulse $V_T$ to an output node k with negative $\delta_k$ for a duration $\Delta t = \gamma|\delta_k|$ correctly updates the output weights with negative $V_j^{\text{int}}$ and negative $\delta_k$, while leaving all other weights fixed.

Now, we can update the weights associated with positive $V_j^{\text{int}}$ and negative $\delta_k$ by simply applying a negative pulse $-V_T$ for the same duration. A quick check shows that now, nodes with negative $V_j^{\text{int}}$ receive no conductance change, and the direction of the voltage drop for the positive $V_j^{\text{int}}$ is flipped. This reverses the direction of the conductance change, updating all weights associated with positive $V_j^{\text{int}}$ in the correct direction and magnitude. This method thus handles the second case, with positive $V_j^{\text{int}}$ and negative $\delta_k$.

To summarize so far, applying a positive pulse $V_T$ followed by a negative pulse $-V_T$, both for duration $\Delta t = \gamma|\delta_k|$, to an output node k with negative $\delta_k$ will correctly update all the weights into that output node, while leaving all other weights fixed. Now, we must show how weights associated with output nodes having positive $\delta_k$ can be updated.

It is helpful to note that moving from output nodes with negative $\delta_k$ to positive $\delta_k$ requires reversing the directions of the updates compared to the negative case. To reverse the directions of these updates, we can reverse the signs of the internal node voltages $V_j^{\text{int}}$ prior to applying the voltage pulses at the output node k. This will reverse the voltage drops which adjust the conductances at nodes $V_j^{\text{int}}$.

Luckily, this is straightforward to do. Prior to applying the positive and negative voltage pulses at the output node k, we can flip the signs of all the input voltages to the network. Since Eq. (2) is an odd function with respect to the input voltages $V_i^{\text{ext}}$, flipping all the input voltages will also flip the signs of all the $V_j^{\text{int}}$'s. This will flip the signs of the conductance changes compared to the negative $\delta_k$ case, while maintaining the correct magnitudes. Thus, we have shown how to update nodes with negative $\delta_k$: simply reverse all input voltages prior to applying a positive pulse $V_T$ followed by a negative pulse $-V_T$, each for a duration $\Delta t = \gamma|\delta_k|$.

To perform this process efficiently, we update all output nodes with negative $\delta_k$ using positive and negative voltage pulses with duration $\Delta t = \gamma|\delta_k|$, then flip the signs of all the inputs and update all the nodes with positive $\delta_k$, again using positive and negative voltage pulses with duration $\Delta t = \gamma|\delta_k|$.

*B. Input Weight Updates*

Now that we have a mechanism for updating the output weights, we must extend the delta rule to the full backpropagation algorithm in order to update the input weights. For the input weights of a conventional linear two-layer network, the correct updates are given by:

$$\Delta W_{ji} = \alpha \left( \sum_k W_{kj} (y_k - T_k) \right) x_i \quad (8)$$

We see that the delta $\delta_k = (y_k - T_k)$ of Eq. (5) has been replaced by a weighted sum of deltas $\left( \sum_k W_{kj} (y_k - T_k) \right)$ into node j. This weighted sum of deltas is node j's *contribution* to the error. It tells us which direction $V_j^{\text{int}}$ should move in order to decrease the error. With this information, we can move $V_j^{\text{int}}$ by adjusting the input weights $W_{ji}$ in order to decrease the error.

Mapping variables in a similar way as we did from Eq. (5) to Eq. (6), we arrive at:

$$\Delta G_{ji} = \alpha \left( \sum_k G_{kj} \left( \sigma(I^{\text{ext}})_k - T_k \right) \right) V_i^{\text{ext}} \quad (9)$$

Note that again, the actual derivation of this equation is more complicated due to the normalization term $G_j$ of Eq. (2), which is not present in standard neural networks. We ignore the normalization term, as in the previous case of the output weights.

We once again have a situation where the correct weight update $\Delta G_{ji}$ is given as a product of two terms: $\delta_j = \left( \sum_k G_{kj} \left( \sigma(I^{\text{ext}})_k - T_k \right) \right)$, and $V_i^{\text{ext}}$. If we can reproduce the $\delta_j$ as a voltage at the internal node j and apply pulses $V_T$ and $-V_T$ to the input electrodes for a duration $\Delta t = \gamma|V_i^{\text{ext}}|$, we can apply the same methodology as the previous section to update the input weights.

Consider what happens when we ground the input nodes of the network and place voltages at the output nodes proportional to their deltas $V_k^{\text{ext}} = \lambda \delta_k$. Now, the voltages at the internal nodes are given by:

$$V_j^{\text{int}} = \frac{\sum_k G_{kj} V_k^{\text{ext}}}{G_j} = \frac{\sum_k G_{kj} \lambda \left( \sigma(I^{\text{ext}})_k - T_k \right)}{G_j} \quad (10)$$

This essentially means that we have successfully reproduced $\delta_j$ as a voltage at internal node j, up to a normalization constant $G_j$ and scaling factor $\lambda$. Following the process above, we can apply voltages $V_T$ and $-V_T$ to the input nodes for a duration proportional to the original input voltage $\Delta t = \gamma|V_i^{\text{ext}}|$ to update the input weights. The voltages applied to the outputs, which are proportional to the deltas $V_k^{\text{ext}} = \lambda \delta_k$ can be flipped or not flipped prior to the application of the pulses to handle both cases of positive and negative $V_i^{\text{ext}}$, ensuring the correct sign of the update.

Even with the undesirable normalization term $G_j$, the sign of $\delta_j$ is guaranteed to be correct, since $G_j$ is always non-negative. As mentioned above, we find that the effect of the normalization term is minimal and decreases with increasing network size. This normalization term and the voltage perturbations of $V_j^{\text{int}}$ from applying the pulses $V_T$ and $-V_T$ are the primary sources of error in our idealized network model. Both these sources of error are accounted for in our network simulations.

Since the above algorithm only iterates over each external node once per training example, we see that it is first order in time O(n) in the number of neurons per layer. This is in contrast with the standard backpropagation algorithm when performed on conventional hardware (e.g. GPUs), which is second order in time O(n$^2$) in the number of neurons per layer, due to the scaling of matrix multiplication.

The O(n) time complexity for backpropagation combined with O(n) spatial scaling and efficient long-range connectivity is unique to the MN3, thus making it very attractive for training ultra-wide deep neural networks.

The pseudo-code for the MN3 training algorithm is shown in Algorithm (1).

---

**Algorithm 1** MN3 Training Algorithm

Given: *S* training samples, $N_1$ input nodes, $N_2$ output nodes, $N = N_1 + N_2$ total nodes, *M* nanowires, input data voltages $V_i^{\text{ext}} \in \left[-\frac{V_T}{2}, \frac{V_T}{2}\right]$, conductances $G \in \mathbb{R}^{N \times M}$, targets $T_1$, $T_2$, …, $T_{N2}$, constant of proportionality $\lambda$, and k' = k – $N_1$ for indexing purposes.

---

1: **for** *s*=1:*S* **do**
2:    $V_{1:N_1}^{\text{ext}}$ = data[*s*]
3:    $V_{N_1+1:N}^{\text{ext}} = 0$        # ground all output nodes
4:    $G_j = \sum_{i=1}^{N} G_{ij}$
5:    $V_j^{\text{int}} = \frac{\sum_{i=1}^{N} G_{ij} V_i^{\text{ext}}}{G_j}$
6:    $I_{k'}^{\text{ext}} = \sum_{j=1}^{M} G_{kj} V_j^{\text{int}}$
7:    **for** *k*=$N_1$+1:*N* **do**
8:       **if** $T_{k'} - I_{k'}^{\text{ext}} < 0$ **do**
9:          Apply $V_i^{\text{ext}}$ at each input node
10:         Apply +$V_T$ to output node k' for $t \propto T_{k'} - I_{k'}^{\text{ext}}$
11:         Apply -$V_T$ to output node k' for $t \propto T_{k'} - I_{k'}^{\text{ext}}$
12:       **if** $T_{k'} - I_{k'}^{\text{ext}} > 0$ **do**
13:          Apply $-V_i^{\text{ext}}$ at each input node
14:          Apply +$V_T$ to output node k' for $t \propto T_{k'} - I_{k'}^{\text{ext}}$
15:          Apply -$V_T$ to output node k' for $t \propto T_{k'} - I_{k'}^{\text{ext}}$
16:    $V^{\text{err}} = \lambda(T - I^{\text{ext}})$     # converts current to voltage
17:    $V_{\text{prev}}^{\text{ext}} = V_{1:N_1}^{\text{ext}}$
18:    $V_{1:N_1}^{\text{ext}} = 0$         # ground all input nodes
19:    **for** *i*=1:$N_1$ **do**
20:       **if** $V_{\text{prev},i}^{\text{ext}} > 0$ **do**
21:          Apply $V_k^{\text{err}}$ at each output node
22:          Apply +$V_T$ to input node *i* for $t \propto |V_{\text{prev},i}^{\text{ext}}|$
23:          Apply -$V_T$ to input node *i* for $t \propto |V_{\text{prev},i}^{\text{ext}}|$
24:       **if** $V_{\text{prev},i}^{\text{ext}} < 0$ **do**
25:          Apply $-V_k^{\text{err}}$ at each output node
26:          Apply +$V_T$ to input node *i* for $t \propto |V_{\text{prev},i}^{\text{ext}}|$
27:          Apply -$V_T$ to input node *i* for $t \propto |V_{\text{prev},i}^{\text{ext}}|$

## V. SIMULATION RESULTS

There are several structural components that are unique to the MN3's architecture and the aforementioned training algorithm that must be considered. We must account for them in the simulations accordingly, and they are as follows:

*Bounded, Non-negative Weights*: Since the weights in the network are represented in hardware by the conductances at each electrode-nanowire junction, the weights are non-negative. If a nanowire does not pass over a given electrode, then the conductance is zero. Further, memristors possess lower and upper bounds on their possible conductances. However, for these simulations, we allow for all weights in the set of non-negative real numbers. The use of non-negative weights does not hinder network performance when activation functions that span both negative and positive values are used as this permits logical inverting of input [39]. See appendix subsection (A.1) for weight initialization details.

*Composition of Linear Layers*: As described above, one forward pass through the MN3 is akin to performing two linear transformations with a normalization term placed between them. This, of course, differs from the standard linear layer in the context of deep learning, where one single linear transformation is performed without any normalization term.

*Noisy Activations and Gradients*: Due to the nature of analog components, the activations and gradients (and weights by extension) possess some degree of noise. There is a significant amount of literature demonstrating that noise injection improves the stability and convergence of networks [40, 41, 42]. We add multiplicative Gaussian noise in the weight update gradients, i.e., $\mathbf{W} = \mathbf{W} + \nabla\mathbf{W} * (1 + N(0, \sigma^2))$, where $\sigma = 0.05$.

*Bounded Activations*: Memristors possess positive and negative threshold voltages $V_{TP}$ and $V_{TN}$, respectively, where any voltage between these two values will not cause switching. These values are often asymmetric ($|V_{TP}| \neq |V_{TN}|$). In the forward pass, we must ensure that all activations are between these two values. One way to do this is to use a function that squashes the input voltages into the acceptable voltage range. This can be achieved by using the tanh function since it maps all real numbers to the range [-1, 1]. In order to determine the range, first assume $V_T = V_{TP} = -V_{TN}$. Next, see that if $V_T / 2$ is placed on an input electrode and a nanowire in contact with this electrode has voltage -$V_T / 2$, then we have $\Delta V = V_T$, the maximum voltage in which switching does not occur. For asymmetric cases, it suffices to scale the voltages by the threshold voltage with the smaller magnitude. Hence, for these simulations, we use the activation function $\sigma(\mathbf{v}) = \min(|V_{TP}|, |V_{TN}|)/ 2 * \tanh(\mathbf{v})$, which, as desired, results in MN3 input voltages $\mathbf{v}$ with $v \in [-\min(|V_{TP}|, |V_{TN}|) / 2, \min(|V_{TP}|, |V_{TN}|) / 2]$ for $v \in \mathbf{v}$. This activation function doubles as both a voltage range restrictor and a nonlinear function. Further, it spans both negative and positive values, enabling higher performance when using non-negative weights [39]. It is important to note that these simulations assume symmetric threshold voltages ($V_{TP} = 2$ V and $V_{TN} = -2$ V). Nonetheless, we also tested asymmetric values and found minimal differences in performance. For example, we simulated $V_{TP} = 0.75$ V and $V_{TN} = -0.5$ V, which are realistic threshold voltages found in hafnium-oxide [43] and produced similar results.

*Sparse Connectivity*: Unlike standard fully connected layers, where each neuron at one layer is connected to every neuron of the next layer, the MN3 is sparsely connected, with sparsity of ~97-98% in each of the two linear layers for the MN3 sizes tested. Specifically, this type of sparsity forms a small-world network [27]. Small-world neural networks at this level of sparsity have been shown to outperform random networks at the same level of sparsity [34].

*No Batch Updates*: A batch size of 1 must be used due to the architecture of the MN3. This is because each electrode can only support one voltage at a time.

There are a few standard deep learning methods that should be employed in software when training a deep MN3 network as well. The two most important techniques are below:

*Summed Outputs*: For tasks with a small number of outputs relative to the number inputs (e.g. 784 inputs neurons and 10 output neurons), our simulations show that the perturbation in internal node voltages caused by pulsing the external nodes with the threshold voltages is relatively large. To mitigate this, we increase the number of output neurons (e.g. from one neuron per class to 10 or 100 neurons per class) and sum over designated groups of neurons that correspond to a specific class. Backpropagation between the loss function and the MN3's output layer is done in software using the machine learning framework PyTorch's autograd functionality and is a routine computation [44].

*Layer Normalization*: Layer normalization uses the mean and variance of all outputs of a layer in a single training step to normalize the activations [45]. We have found that one should use layer normalization on the outputs of each MN3 as it improves convergence greatly. However, one need not use it on the final layer if using the softmax activation function.

To determine the effectiveness of both the MN3 architecture and the training algorithm, we tested a variety of models. We tested both sequential and non-sequential datasets to show the MN3's ability to support both recurrent networks and feedforward networks, respectively.

### A. MNIST

We used the MNIST handwritten digit dataset to test for simple image recognition ability [46]. This dataset is composed of 28 × 28 pixel grayscale images, where each image shows a handwritten digit from 0 to 9. There are 60,000 training images and 10,000 testing images. Other than normalizing the input data to the standard normal distribution, no data augmentation was used. We tested one, two, three, and six-layer standard multi-layer perceptrons (MLPs). Standard backpropagation and stochastic gradient descent (SGD) were used. Momentum was not used. Then, we trained one, two, three, and six-layer MLPs with the constraints listed above and the algorithm proposed in this paper. The hyperbolic tangent activation function was used after each layer (other than the nanowire) prior to the final layer, where the softmax activation function was used. The results are outlined in Table II. The error gap between the standard MLPs and MN3 MLPs is ~0-2%. The test errors for the 6-layer MLPs show that for increasing depth, the MN3 achieves comparable performance to standard approaches.

*B. pMNIST*

We will now test the MN3's ability to function as a recurrent neural network (RNN), specifically as a long-short term memory (LSTM) network. To do this, we will use the pMNIST, or permuted MNIST, dataset, where pixels are randomly permuted, and the same permutation is used for each training and testing sample [47]. We perform row-by-row training and testing, where a row of the image is fed into the network at each time step. The implementation follows a standard LSTM implementation, except an MN3 layer followed by layer normalization replaces every fully connected layer. For the classification layer, which is only called once per sample (and not at each time step within a sample), a standard software-based fully connected layer is used. This is reasonable as the classification layer only makes up a very small fraction of the total number of parameters and computation time required in an LSTM. The results are listed in Table III.

TABLE III
pMNIST classification errors for a standard LSTM trained with backpropagation and SGD and simulated MN3 LSTM trained with the algorithm described in this paper.

| Method | Error % |
|---|---|
| Standard LSTM (784 — 1000 all hidden layers — 10) | 2.29% |
| MN3 LSTM (784 — 1000 all hidden layers, 2048 nanowires — 10) | 3.67% |

It is clear that the standard LSTM outperforms the MN3 LSTM by less than 2%. There is strong evidence that increasing the width and depth of LSTMs can greatly improve final test accuracy [32, 34]. Specifically, in [34], Gray et al. obtain state-of-the-art results in several sequential modeling tasks using an LSTM with ~97% sparsity and a state size of 18432.

## VI. FUTURE WORK

There are multiple orthogonal developments left to be made in both harnessing the MN3 in the context of deep learning and simulation. We list relevant research directions here.

TABLE II
MNIST classification errors for standard MLPs trained with backpropagation and SGD and simulated MN3s trained with the algorithm described in this paper. Each network has 784 input units (28 × 28 pixels). The number of units in each hidden layer is shown. For the standard MLPs, the output layers contain 10 output units – one for each class. For the MN3 MLPs, the output layers contain 100 units, and these outputs are then summed over groups of 10 to form 10 classes. This is done to mitigate the effect of voltage perturbations as discussed above.

| Standard MLPs + LN | # params | Error | Error | # params | MN3 MLPs |
|---|---|---|---|---|---|
| 2-layer (784 — 1000 — 10) | 795,794 | 1.62% | 3.9% | 118,128 | 2-layer (784 — 1000 — 100, 2048 nanowires) |
| 3-layer (784 — 1000 — 1000 — 10) | 1,796,794 | 1.49% | 2.41% | 200,048 | 3-layer (784 — 1000 — 1000 — 100, 2048 nanowires) |
| 6-layer (784 — 2500 — 2000 — 1500 — 1000 — 500 — 10) | 11,973,294 | 1.48% | 1.61% | 1,293,844 | 6-layer (784 — 2500 — 2000 — 1500 — 1000 — 500 — 100, 4096 nanowires) |

*Complex Networks*: Compared to neural networks that are implemented on conventional hardware, such as GPUs and crossbar-based neuromorphic hardware, networks created using the MN3 can be much larger due to built-in sparsity and by removing the neuron-synapse connectivity from the CMOS layer. Further, as is typical with neuromorphic hardware, training speed scales linearly with additional neurons and not quadratically, as is the case with standard digital approaches. This opens up the possibilities for an untapped class of ultra-large, trainable networks. One application area of interest is Transformer models, which are mainly composed of large fully connected layers [31]. We note that one can achieve state-of-the-art results in language tasks simply by scaling up the dimension of the feedforward layer in transformer models [53].

*Alternative Algorithms*: The training algorithm presented in this paper is not the only acceptable training algorithm that may be implemented using the MN3. For example, with some minor changes, one could implement Equilibrium Propagation, a learning framework where only one type of neural computation is needed for both the forward and backward pass, using the MN3 [48]. We have shown that Equilibrium Propagation provides a better approximation to the gradients and is currently under development. Further, there is great promise in reframing the algebra behind the MN3, an electrical network, in terms of tensor networks, which provide a more natural framework for performing both linear and nonlinear transformations in electrical networks. G. Kron gives a strong theoretical foundation for this in his *Tensor Analysis of Networks* [49].

*Electrode Topologies*: In the simulations presented in this paper, we mapped data randomly to the set of MN3 electrodes; however, we hypothesize that manipulating the MN3's electrode topology can increase performance. For example, with image data, we posit that mapping input pixels onto the input electrodes without compromising the natural locality of the image and interspersing output electrodes evenly between input electrodes will produce a convolution-like operation. Further, it is an extremely simple change in hardware. We are interested in mapping other types of data to appropriate electrode topologies as well. Additionally, MN3s with a range of electrode sizes can be used to produce scale-free networks with a diverse degree distribution.

*Bayesian Framework*: As is common with analog signals, memristors operate in the low-precision domain [50]. Unlike low-precision in digital hardware, where $N$-bit precision implies there are $2^N$ unique values, an analog component with $N$-bit precision can assume a continuous range of values, however the precision is limited due to stochasticity that is intrinsic to the device. For this reason, it may be appropriate to adopt a Bayesian framework, where the distributions around each possible value assumed by a memristor are taken into consideration. That is, instead of learning a matrix of point-estimates for our weight parameters, we learn a matrix of distributions. A detailed explanation of applying Bayesian learning to neural networks is given in [51].

*Realistic Memristor Model*: While the simple linear threshold model used in this paper is sufficient in deriving the backpropagation algorithm, there exist more detailed nuances specific to particular memristive systems that should be encapsulated in later iterations. Most importantly, variability in the memristor threshold and asymmetric nonlinearity in the update should be taken into consideration. These realistic models — along with compatible algorithms — are currently under development.

## VII. Conclusion

We have presented a gradient-based training algorithm for the Memristive Nanowire Neural Network (MN3), a novel neuromorphic chip architecture, and shown its applications in the context of deep learning. Specifically, we showed how the MN3 could supplant large, fully connected layers in deep networks. We showed that the MN3 can achieve comparable results on the MNIST and pMNIST datasets to standard software-based networks. We have open-sourced our code as well as a few tutorials so that those in the deep learning community can experiment with the MN3 and its applications. Importantly, we note that the algorithm presented is not the only algorithm that the MN3 can support. This paper serves as a template for readers to envision alternative algorithms to run on the presented (or modified) hardware and application areas suited for ultra-wide networks.

## VIII. APPENDIX

### A.1. MN3 Weight Initialization

We initialize weights by modifying the Xavier initialization heuristic [52]. Xavier initialization states that for maximizing stability and convergence, one should initialize weight matrix $\mathbf{W}$ by

$$\mathbf{W} \sim U\left[-\frac{\sqrt{6}}{\sqrt{n_1 + n_2}}, \frac{\sqrt{6}}{\sqrt{n_1 + n_2}}\right]^{n_1 \times n_2}, \quad (9)$$

where $U[a, b]$ is the uniform distribution on $[a, b]$ and $n_1$ and $n_2$ are the number of units in the layers before and after weight matrix $\mathbf{W}$, respectively. Since the MN3's weights are non-negative, we instead initialize $\mathbf{W}$ by

$$\mathbf{W} \sim U\left[0, \frac{2\sqrt{6}}{\sqrt{n_1 + n_2}}\right]^{n_1 \times n_2}. \quad (10)$$

We choose this initialization as it preserves the variance of standard Xavier weight initialization – namely, immediately following initialization, we have $\text{Var}(\mathbf{W}) = 2 / (n_1 + n_2)$.

### A.2. MN3 Forward Pass

Consider data $D = \{(\mathbf{x}_1, y_1), \ldots, (\mathbf{x}_n, y_n)\}$, where $\mathbf{x}_i \in \mathbb{R}^p$ and $y_i \in \mathbb{R}$, and an MN3 with weights $\mathbf{W}_{\text{in}} \in \mathbb{R}^{p \times q}$ and $\mathbf{W}_{\text{out}} \in \mathbb{R}^{q \times r}$ initialized using (10). For $\mathbf{x} \in \{\mathbf{x}_1, \mathbf{x}_2, \ldots, \mathbf{x}_n\}$ and $y \in \{y_1, y_2, \ldots, y_n\}$, the forward pass is given analytically by the following series of equations (excluding stochasticity):

$$\begin{aligned} \mathbf{h}_{\text{in}} &= \mathbf{W}^T_{\text{in}} \mathbf{x}, \\ \mathbf{h} &= \mathbf{h}_{\text{in}} / \mathbf{G}, \qquad (11) \\ \mathbf{y}_{\text{in}} &= \mathbf{W}^T_{\text{out}} \mathbf{h}, \end{aligned}$$

where

$$\mathbf{G} = \sum_{i=1}^{p} \mathbf{W}_{\text{in}; i,:} + \sum_{k=1}^{r} \mathbf{W}_{\text{out}; :,k}. \quad (12)$$

Note that the three calculations in (11) are done implicitly in the MN3, where input data $\mathbf{x}$ is fed into the MN3 as voltages and the result of the two matrix-multiplies and normalization step is read out as currents.

### A.3. MN3 Deltas

To form a complete network, we append two steps to the forward pass presented above:

$$\begin{aligned} \mathbf{h}_{\text{in}} &= \mathbf{W}_{\text{in}} \mathbf{x}, \\ \mathbf{h} &= \mathbf{h}_{\text{in}} / \mathbf{G}, \\ \mathbf{y}_{\text{in}} &= \mathbf{W}_{\text{out}} \mathbf{h}, \qquad (13) \\ \mathbf{y}_{\text{out}} &= \sigma(\mathbf{y}_{\text{in}}), \\ L &= f(\mathbf{y}_{\text{out}}, y), \end{aligned}$$

where $\sigma : \mathbb{R}^r \to \mathbb{R}^r$ is the softmax activation function and $f : (\mathbb{R}^r, \mathbb{R}) \to \mathbb{R}$ is the single-label cross-entropy loss function. Note that the last two calculations are done in software.

The computational graph of the forward pass and relevant partial derivatives are shown in Fig. 5. The exact calculations for each partial derivative in the above computational graph are shown below. We exclude $\partial L / \partial \mathbf{y}_{\text{out}}$ as it is well known that $\partial L / \partial \mathbf{y}_{\text{in}} = \mathbf{y}_{\text{out}} - \mathbf{t}$, where $\mathbf{t}$ is the one-hot encoded form of target $y$, when the softmax activation function and cross-entropy loss function are used.

$$\begin{aligned} \frac{\partial L}{\partial \mathbf{y}_{\text{in}}} &= \mathbf{y}_{\text{out}} - \mathbf{t} & \frac{\partial \mathbf{y}_{\text{in}}}{\partial \mathbf{h}} &= \mathbf{W}^T_{\text{out}} & \frac{\partial \mathbf{h}}{\partial \mathbf{h}_{\text{in}}} &= \text{diag}\left(\mathbf{G}^{-1}\right) \\ \frac{\partial \mathbf{h}_{\text{in}}}{\partial \mathbf{x}} &= \mathbf{W}^T_{\text{in}} & \frac{\partial \mathbf{y}_{\text{in}}}{\partial \mathbf{W}_{\text{out}; jk}} &= h_j \mathbf{e}_k & \frac{\partial \mathbf{h}}{\partial \mathbf{G}} &= \text{diag}\left(-\mathbf{h}_{\text{in}} \cdot \mathbf{G}^{-2}\right) \\ \frac{\partial \mathbf{h}_{\text{in}}}{\partial \mathbf{W}_{\text{in}; ij}} &= x_i \mathbf{e}_j & \frac{\partial \mathbf{G}}{\partial \mathbf{W}_{\text{out}; jk}} &= \mathbf{e}_j & \frac{\partial \mathbf{G}}{\partial \mathbf{W}_{\text{in}; ij}} &= \mathbf{e}_j \end{aligned}$$
(14)

*A.4. MN3 Analytic Gradients*

Using the information above, we can calculate the analytic gradients of $\mathbf{W}_{in}$ and $\mathbf{W}_{out}$. For $\mathbf{W}_{out}$, we have

$$\frac{\partial L}{\partial \mathbf{W}_{out;jk}} = \frac{\partial \mathbf{y}_{in}}{\partial \mathbf{W}_{out;jk}} \cdot \frac{\partial L}{\partial \mathbf{y}_{in}} + \frac{\partial \mathbf{G}}{\partial \mathbf{W}_{out;jk}} \cdot \frac{\partial \mathbf{h}}{\partial \mathbf{G}} \cdot \frac{\partial \mathbf{y}_{in}}{\partial \mathbf{h}} \cdot \frac{\partial L}{\partial \mathbf{y}_{in}}$$

$$= \left( \frac{\partial \mathbf{y}_{in}}{\partial \mathbf{W}_{out;jk}} + \frac{\partial \mathbf{G}}{\partial \mathbf{W}_{out;jk}} \cdot \frac{\partial \mathbf{h}}{\partial \mathbf{G}} \cdot \frac{\partial \mathbf{y}_{in}}{\partial \mathbf{h}} \right) \left( \frac{\partial L}{\partial \mathbf{y}_{in}} \right)$$

$$= \left( \mathbf{h}_j \mathbf{e}_k + \mathbf{e}_j \operatorname{diag}\left(-\mathbf{h}_{in} \cdot \mathbf{G}^{-2}\right) \mathbf{W}_{out}^T \right) (\mathbf{y}_{out} - \mathbf{t}),$$

(15),

where $\mathbf{e}_a$ is the standard basis. For $\mathbf{W}_{in}$, we have

$$\frac{\partial L}{\partial \mathbf{W}_{in;ij}} = \frac{\partial \mathbf{h}_{in}}{\partial \mathbf{W}_{in;ij}} \cdot \frac{\partial \mathbf{h}}{\partial \mathbf{h}_{in}} \cdot \frac{\partial \mathbf{y}_{in}}{\partial \mathbf{h}} \cdot \frac{\partial L}{\partial \mathbf{y}_{in}} + \frac{\partial \mathbf{G}}{\partial \mathbf{W}_{in;ij}} \cdot \frac{\partial \mathbf{h}}{\partial \mathbf{G}} \cdot \frac{\partial \mathbf{y}_{in}}{\partial \mathbf{h}} \cdot \frac{\partial L}{\partial \mathbf{y}_{in}}$$

$$= \left( \frac{\partial \mathbf{h}_{in}}{\partial \mathbf{W}_{in;ij}} \cdot \frac{\partial \mathbf{h}}{\partial \mathbf{h}_{in}} + \frac{\partial \mathbf{G}}{\partial \mathbf{W}_{in;ij}} \cdot \frac{\partial \mathbf{h}}{\partial \mathbf{G}} \right) \left( \frac{\partial \mathbf{y}_{in}}{\partial \mathbf{h}} \cdot \frac{\partial L}{\partial \mathbf{y}_{in}} \right)$$

$$= \left( \mathbf{x}_i \mathbf{e}_j \operatorname{diag}\left(\mathbf{G}^{-1}\right) + \mathbf{e}_j \operatorname{diag}\left(-\mathbf{h}_{in} \cdot \mathbf{G}^{-2}\right) \right) \left( \mathbf{W}_{out}^T \mathbf{y}_{out} - \mathbf{t} \right).$$

(16).

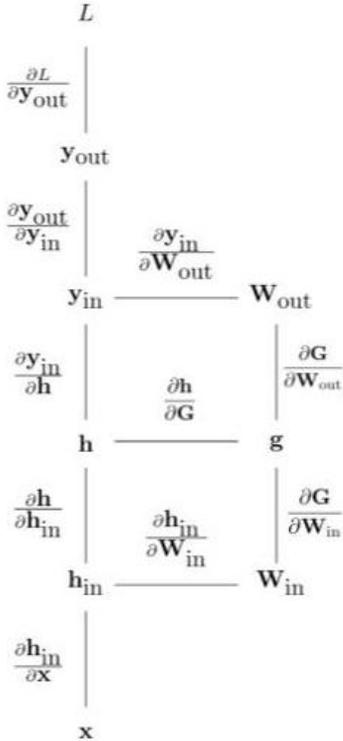

**Fig. 5.** Computational graph of the MN3 with softmax activation function and cross-entropy loss function; dependencies are implied by the partial derivatives.

We note that the training algorithm presented in this paper provides approximate gradients, taking into account only the first additive term in both (15) and (16). This is because the second additive term is small in comparison due to the $\mathbf{g}^{-2}$ term that tends to the zero vector for increasing weight matrix dimensions. Note that $G_j = \sum_{i=1}^{N} G_{ij}$ is the normalization term for internal node j, which grows larger for increasing network size.

ACKNOWLEDGMENT

The authors would like to thank Richard Cai for his contributions to the MN3 Python Library. The authors would also like to thank Laura Suarez for her role in the discussions which inspired the algorithm presented in this paper. This work was supported by Rain Neuromorphics, Inc. JDK and JCN are co-owners and RDP is an employee of Rain Neuromorphics, Inc. JCN acknowledges the support of the US National Science Foundation under Grant No. ECCS-1709641.